\documentclass[preprints,review,accept,pdftex,oneauthor]{Definitions/mdpi} 
\firstpage{1} 
\pubvolume{1}
\issuenum{1}
\articlenumber{0}
\pubyear{2025}
\copyrightyear{2025}
\datereceived{ } 
\daterevised{ } % Comment out if no revised date
\dateaccepted{ } 
\datepublished{ } 
\hreflink{https://doi.org/} % If needed use \linebreak
\Title{Symbiotic stars in the era of modern ground- and space-based surveys}

\TitleCitation{Symbiotic stars in the era of modern ground- and space-based surveys}

\Author{Jaroslav Merc $^{1,2}$\orcidA{}}

\AuthorNames{Jaroslav Merc}

\isAPAStyle{%
       \AuthorCitation{Merc, J.}
         }{%
        \isChicagoStyle{%
        \AuthorCitation{Lastname, Firstname, Firstname Lastname, and Firstname Lastname.}
        }{
        \AuthorCitation{Merc, J.}
        }
}

\address{%
$^{1}$ \quad Astronomical Institute of Charles University, V Hole\v{s}ovi\v{c}k{\'a}ch 2, Prague, 18000, Czech Republic; jaroslav.merc@mff.cuni.cz\\
$^{2}$ \quad Instituto de Astrof\'isica de Canarias, Calle Vía Láctea, s/n, E-38205 La Laguna, Tenerife, Spain}

\corres{Correspondence: jaroslav.merc@mff.cuni.cz}

\abstract{Symbiotic stars, interacting binaries composed of a cool giant and a hot compact companion, exhibit complex variability across the electromagnetic spectrum. Over the past decades, large-scale photometric and spectroscopic surveys from ground- and space-based observatories have significantly advanced their discovery and characterization. These datasets have transformed the search for new symbiotic candidates, providing extensive time-domain information crucial for their classification and analysis. This review highlights recent observational results that have expanded the known population of symbiotic stars, refined classification criteria, and enhanced our understanding of their variability. Despite these advances, fundamental questions remain regarding their long-term evolution, mass transfer and accretion processes, or their potential role as progenitors of Type Ia supernovae. With ongoing and upcoming surveys, the coming years promise new discoveries and a more comprehensive picture of these intriguing interacting systems.}


\begin{document}

%%%%%%%%%%%%%%%%%%%%%%%%%%%%%%%%%%%%%%%%%%

\section{Introduction}

Given that a significant fraction of stars do not evolve in isolation, unlike our Sun, understanding the interactions within binary (and multiple) systems is essential for predicting their evolution and ultimate fate. These interactions drive some of the most energetic and complex astrophysical phenomena, including kilonovae, gamma-ray bursts, gravitational wave emission from compact binary mergers, Type Ia supernovae, or nova outbursts, and create the most exotic remnants like Thorne-Żytkow objects, or double-degenerate binaries \citep[see, e.g.,][]{2017ApJS..230...15M,2023ASPC..534..275O,2023hxga.book..129B}.

Symbiotic stars represent a particularly intriguing subgroup within the diverse binary population. These systems, consisting of a compact accretor, typically a white dwarf or neutron star, and a cool giant donor \citep{2012BaltA..21....5M,2019arXiv190901389M}, exhibit a wide range of observable effects driven by mass transfer and accretion. These include the formation of accretion disks, collimated jets, thermonuclear outbursts, as well as phenomena associated with the surrounding dense, photoionized nebula, colliding stellar winds, or the production and destruction of dust. The symbiotic interaction manifests across an extensive range of wavelengths, from gamma rays and X-rays to infrared and radio, and on timescales spanning minutes to decades. Many of the aforementioned astrophysical processes are shared with other types of stellar systems. Studying symbiotic stars is thus crucial not only for understanding binary interactions but also for their broader implications in stellar evolution, nucleosynthesis, and even extragalactic studies, as they have been proposed as possible progenitors of Type Ia supernovae \citep{1992ApJ...397L..87M, 1999ApJ...522..487H,2013IAUS..281..162M, 2016A&A...588A..88M,2019MNRAS.485.5468I,2023RAA....23h2001L,2025A&ARv..33....1R}. Their evolutionary connection to post-common-envelope systems and double white dwarf binaries makes them relevant for understanding the population of gravitational wave sources detectable by the Laser Interferometer Space Antenna (LISA), while their accretion-powered activity provides valuable analogies to other interacting systems, from X-ray binaries to active galactic nuclei.

Despite their importance, symbiotic stars have long remained a relatively small and specialized research field, with only about 300 confirmed systems identified to date \citep{2019RNAAS...3...28M,2019AN....340..598M,2019ApJS..240...21A}. Many fundamental questions regarding their formation, evolution, and the processes governing mass transfer and accretion remain unresolved. These questions have become increasingly pressing as modern ground- and space-based surveys begin to unveil new symbiotic candidates and provide large-scale, time-domain datasets to study their variability. This resurgence of interest in symbiotic binaries is evidenced by the recent dedicated conference \textit{'Symbiotic stars, weird novae, and related embarrassing binaries'} \citep{2024NatAs...8.1504M}, which gathered the community for the first time in over a decade, highlighting both the growing observational potential and the need for a coordinated effort to fully exploit these new data. In this manuscript, I review recent advances in the search for and characterization of symbiotic binaries driven by the availability of these observational datasets.

\subsection{On the symbiotic definition}
Before discussing recent observational results, I first briefly outline the definition of symbiotic stars, which has evolved over decades through both observational and theoretical advancements. Initially, these objects were identified by their peculiar optical spectra, resembling those of cool red giants but featuring strong emission lines and a faint blue continuum. The coexistence of low-temperature features (e.g., TiO bands) and high-temperature indicators (e.g., the He\,{\sc ii} 4686 \AA{} emission line) led to their classification as stars with "combination spectra"\footnote{Not to be confused with "composite spectra" stars, a term introduced by Annie Jump Cannon at Harvard Observatory to describe spectra composed of two nearly normal separate stars.}. The term "symbiotic stars" was later introduced by Paul W. Merrill during an American Astronomical Society meeting at Yerkes Observatory in 1941 \citep{1958LIACo...8..436M}.

Early models considered single-star explanations for their properties, but the evidence increasingly favored binarity as the only viable explanation \citep[for a historical review, see][]{1986syst.book.....K}. It is now well established that symbiotic stars are interacting binaries. The donor is an evolved red giant, either on the red giant branch (RGB) or asymptotic giant branch (AGB). The accretor is a compact object, either a white dwarf or a neutron star\footnote{The existence of higher-mass counterparts to symbiotic binaries with black hole accretors has been theoretically predicted \citep[e.g.,][]{2024ApJ...977...95D}, but none have been confirmed to date.}, typically contributing little or no flux in the optical. Some well-known systems were initially thought to contain main-sequence accretors undergoing high accretion, but subsequent observations confirmed white dwarfs as the true accretors. While interacting red giant binaries with main-sequence companions certainly exist, they belong to a different evolutionary class. In symbiotic stars, both components are evolved: the donor is a red giant, and the accretor is already in a late evolutionary stage as a white dwarf or neutron star. This long evolutionary history suggests possible past interactions that may have shaped the current state of the system. I also argue against referring to red giant–main-sequence interacting binaries (systems like SS Lep) as "pre-symbiotic stars" \citep[e.g.,][]{2019arXiv190901389M}, as their evolutionary paths may not necessarily lead to a classical symbiotic binary. For example, if they undergo common-envelope evolution, they could instead become cataclysmic variables with significantly shorter orbital periods.

Based on the above, a symbiotic star must contain a cool red giant (identified through absorption features in the optical or infrared) interacting with a compact companion. While the exact degree of interaction required for classification, often expressed in terms of accretion rate, remains debated, it must lead to observable manifestations at some wavelength \citep[e.g.,][]{2013A&A...559A...6L,2016MNRAS.461L...1M}.

If the accretion rate is high enough to sustain stable surface nuclear burning on the white dwarf (so-called shell-burning symbiotic stars), the companion becomes hot and luminous, ionizing a large portion of the wind of the giant. These systems exhibit a strong nebular continuum and rich emission-line spectra (see Fig. \ref{fig:spectra}). Traditional classification criteria are particularly well suited for identifying such binaries \citep[e.g.,][]{1986syst.book.....K,2000A&AS..146..407B,2013MNRAS.432.3186M}, as they require the presence of H\,{\sc i}, He\,{\sc i}, and highly ionized emission lines (e.g., [O\,{\sc iii}], He\,{\sc ii}, or [Fe\,{\sc vi}], with ionization potentials above 35 eV). Approximately half of all known symbiotic stars also display Raman-scattered O\,{\sc vi} emission lines at 6830\,\AA{} and 7088\,\AA{} \citep[][]{2019ApJS..240...21A,2019RNAAS...3...28M,2019AN....340..598M}. These lines are exclusive to symbiotics, as their formation requires very specific conditions \citep[][]{1989A&A...211L..31S}.

By contrast, in accreting-only symbiotic stars (where the hot component luminosity is powered solely by accretion), the optical spectrum is dominated by the red giant, with very weak or no emission lines. In these cases, evidence of interaction must be sought at other wavelengths, for example, UV observations can reveal flux excess, short-term flickering, or strong emission lines, while X-ray observations often detect hard emission \citep[e.g.,][]{2013A&A...559A...6L,2016MNRAS.461L...1M,2019arXiv190901389M}.

\subsection{Symbiotic zoo}
{It is not only the strength of the emission lines, as discussed in the previous section, that distinguishes individual symbiotic systems from one another. The currently confirmed sample of symbiotic stars in the \textit{New Online Database of Symbiotic Variables} \citep{2019RNAAS...3...28M,2019AN....340..598M} includes 284 systems in the Milky Way and 71 in external galaxies such as the Large and Small Magellanic Clouds, IC 10, M31, M33, NGC 185, NGC 205, and NGC 6822. In addition, there are hundreds of unconfirmed candidates across the Milky Way and additional 15 galaxies\footnote{For current numbers, refer to the online database at \url{https://sirrah.troja.mff.cuni.cz/~merc/nodsv/}.}.}

{As mentioned earlier, the donor stars in these systems are either on the RGB or AGB. While most known symbiotics host M-type giants, some systems feature K-type stars (the so-called yellow symbiotics), S-type stars, or even carbon stars. The latter group is rare in the Milky Way, with only ten confirmed members, but appears more commonly in the Magellanic Clouds, with the lower overall metallicity.}

\begin{figure}[H]
\isPreprints{\centering}{} % Only used for preprints
\includegraphics[width=13.4 cm]{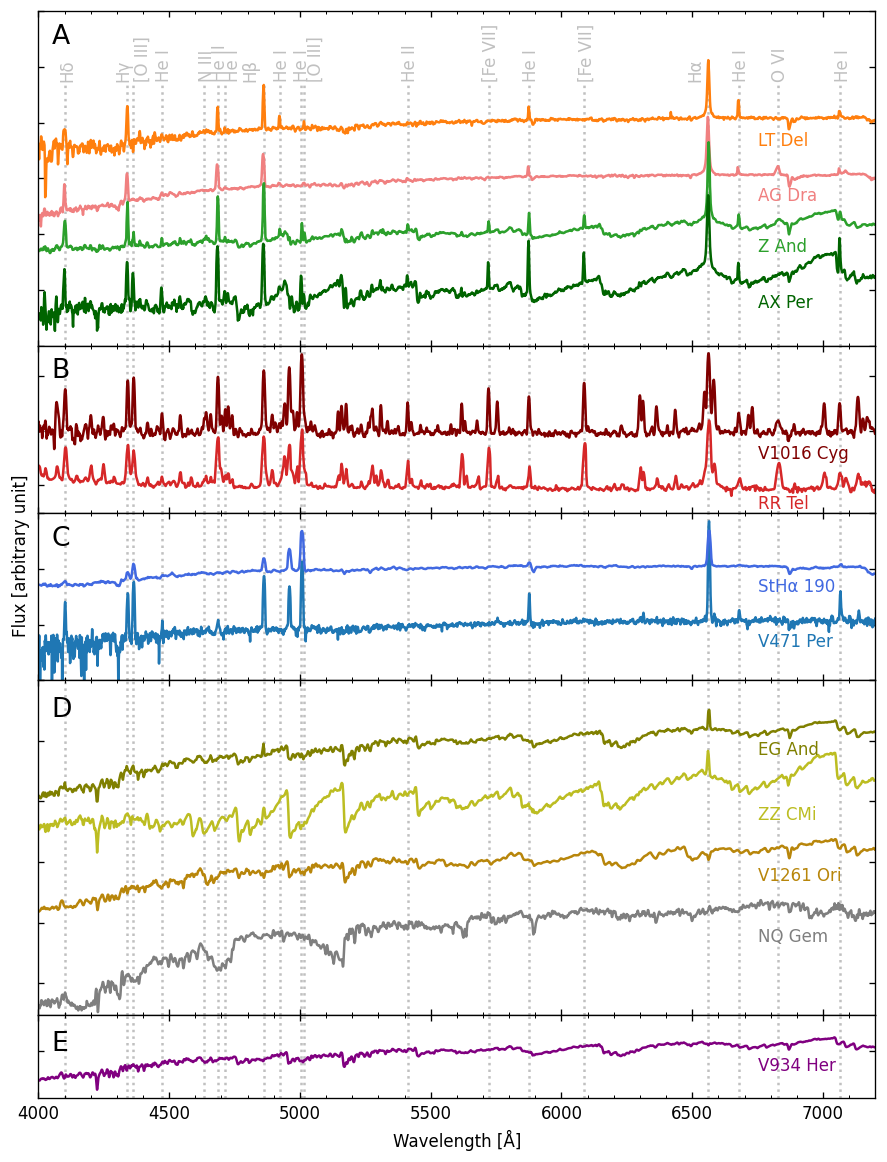}
\caption{Spectra of symbiotic binaries. \textbf{A:} Quiescent spectra of shell-burning S-type ("stellar" infrared type) symbiotic systems: LT Del and AG Dra (yellow symbiotic stars with K-giant donors), and Z And and AX Per (with M-giant donors). \textbf{B:} Quiescent spectra of shell-burning D-type ("dusty") symbiotic systems, V1016 Cyg and RR Tel. \textbf{C:} Quiescent spectra of shell-burning D'-type symbiotic stars, StH$\alpha$\,190 and V471 Per. \textbf{D:} Spectra of accreting-only white dwarf symbiotic stars: EG And and ZZ CMi (with M-giant donors), V1261 Ori (with an S-type giant), and NQ Gem (with a carbon-rich giant). \textbf{E:} Spectrum of the accreting-only symbiotic system V934 Her, which hosts a neutron star accretor. Vertical dotted lines indicate the positions of prominent emission lines commonly observed in shell-burning S-type symbiotic stars. The figure adapted from \citep{2022PhDT........34M}. Spectroscopic data shown were obtained from the ARAS database \citep[][]{2019CoSka..49..217T}.\label{fig:spectra}}
\end{figure}

{The cool giant dominates the near-infrared part of the spectrum. Depending on the evolutionary stage of the giant, the infrared emission varies. In so-called "stellar" infrared-type (S-type) symbiotics, the near-infrared emission is dominated by the photosphere of the giant, typically peaking at $\sim$1.0–1.1 $\mu$m \citep{1995MNRAS.273..517I, 2019ApJS..240...21A}. In contrast, more evolved systems, often showing Mira pulsations, exhibit significant warm dust emission ($\sim$700–1\,000 K; \citep[e.g.,][]{1982ASSL...95...27A}) that shifts the spectral energy distribution peak to $\sim$2–2.5 $\mu$m \citep{2019ApJS..240...21A}. These are referred to as "dusty" (D-type) symbiotics. Some D-types even exhibit two dust components \citep{2010MNRAS.402.2075A}, corresponding to two shells with distinct temperatures. In D-type symbiotic stars, the optical spectrum may lack any trace of the giant companion and resemble that of a planetary nebula (see Fig.~\ref{fig:spectra}).}

{A small group of dusty symbiotics hosting G- or K-type giants show even colder dust components, with SEDs peaking at $\sim$20–30 $\mu$m. These systems are designated as D'-type, following the nomenclature introduced by \citet{1982ASSL...95...27A}. Overall, the symbiotic star population is dominated by S-types (nearly 80\%), with about 15\% classified as D-types and only $\sim$3\% as D'-types.}

{While the cool component dominates the near-infrared and red optical spectrum, the appearance of a symbiotic star at shorter wavelengths (optical, UV, and X-ray) is largely governed by the hot companion. Whether or not the white dwarf is undergoing shell burning depends on several factors, including the mass-loss rate of the donor and the mass of the white dwarf \citep[see, e.g.,][]{2013ApJ...777..136W}. The resulting temperature and luminosity of the hot component drive strong emission lines via photoionization of the surrounding nebula. However, even within a single group, the optical spectra can vary significantly. Among confirmed symbiotic stars in the Milky Way listed in the \textit{New Online Database of Symbiotic Variables}, 67\%, 80\%, 43\%, and 40\% exhibit [O\,{\sc iii}], He\,{\sc ii}, [Fe\,{\sc vi}], and O\,{\sc vi} lines, respectively. Notably, 22\% of systems show He\,{\sc ii} without [O\,{\sc iii}], while virtually none display O\,{\sc vi} without He\,{\sc ii}, confirming the diagnostic value of the He\,{\sc ii} line in verifying O\,{\sc vi} detections, as already suggested, for example,  by \citet{2019ApJS..240...21A}.}

{Approximately 20\% of confirmed Galactic symbiotic stars have been detected in X-rays. It is now clear that their X-ray emission arises from different physical mechanisms across systems. The classification scheme introduced by \citet{1997A&A...319..201M} and expanded by \citet{2013A&A...559A...6L} identifies four primary X-ray types: $\alpha$, $\beta$, $\gamma$, and $\delta$, with some sources showing the combination $\beta$/$\delta$ spectra \citep[see also][]{2019AN....340..598M}. The most common type is $\beta$ (around one third of X-ray symbiotics), characterized by thermal emission below 2.4 keV, likely originating from colliding stellar winds. $\delta$-type systems show harder thermal emission (above 2.4 keV), probably produced in the boundary layer between the accretion disk and the white dwarf. About 20\% of X-ray-detected systems fall into this category, with a similar fraction exhibiting combined $\beta$/$\delta$ features. It is likely that additional symbiotic stars emit X-rays but remain undetected due to current instrument sensitivity limits.}

{Beyond their spectral properties, symbiotic stars also show diverse light curves, which can vary even within the same system depending on the wavelength. In some, the variability is dominated by the cool component, either through semiregular or Mira-type pulsations. In others, orbital variability, such as reflection effects or ellipsoidal modulation, prevails (see Fig.~\ref{fig:lcs}). Approximately one-quarter of all confirmed symbiotic stars in the Milky Way have been observed in at least one outburst.}

\section{Search for symbiotic stars}
The currently known population of symbiotic stars in the Milky Way \citep[$\sim$300;][]{2019RNAAS...3...28M,2019AN....340..598M,2019ApJS..240...21A} falls far short of theoretical predictions, which range from a few thousand to several hundred thousand systems \citep{1984PASA....5..369A,1993ApJ...407L..81K,2003ASPC..303..539M,2006MNRAS.372.1389L,2025arXiv250402090L}. While some symbiotic stars have been (and continue to be) discovered serendipitously, often due to their peculiar photometric behavior, particularly outburst activity, the majority of discoveries in recent years have been enabled by large-scale surveys. Even those initially noticed due to their variability are now predominantly identified through alerts from all-sky photometric surveys (see below).

Historically, many symbiotic stars were first identified through objective prism photographic surveys \citep[e.g.,][]{1973ApJ...185..899S,1978MNRAS.184..601A}, where their spectra appeared similar to those of red giants but featured additional emission lines from both low- and high-ionization species, many of which were unidentified in the early years of research in this field (see the typical symbiotic spectra in Fig. \ref{fig:spectra}). Not a negligible fraction of objects now classified as symbiotic stars were also initially misclassified as planetary nebulae due to their strong emission lines. Their true nature became apparent only later, with the advent of infrared observations, which revealed the presence of cool giant companions \citep[e.g.,][]{1973MNRAS.161..145A,1974MNRAS.168....1A,1974MNRAS.167..337A,1975MNRAS.171..171W}.

\subsection{Emission-line candidates from the narrow-band photometric surveys}
Over the past two decades, new datasets have significantly expanded the known population of symbiotic stars. Many candidates have been identified thanks to the H$\alpha$ photometric surveys covering large regions of the sky. Since symbiotic stars are strong H$\alpha$ emitters (at least the shell-burning ones), they stand out in such datasets, though they appear alongside other emission-line objects such as T Tauri stars and Be stars. Therefore, spectroscopic follow-up is always necessary to confirm their symbiotic nature. Tens of new symbiotic stars and an even larger number of candidates have been discovered thanks to the INT Photometric H$\alpha$ Survey \citep[IPHAS;][]{2005MNRAS.362..753D}, as reported by \citet{2008A&A...480..409C,2010A&A...509A..41C} and \citet{2014A&A...567A..49R} and the AAO/UKST SuperCOSMOS H$\alpha$ Survey \citep[SHS;][]{2005MNRAS.362..689P}, see \citet{2013MNRAS.432.3186M} and \citet{2014MNRAS.440.1410M}. More recently, \citet{2019MNRAS.483.5077A,2021MNRAS.502.2513A} utilized the VST/OmegaCAM Photometric H$\alpha$ Survey \citep[VPHAS+;][]{2014MNRAS.440.2036D}, in combination with additional infrared selection criteria, to identify new candidates.   

Similar photometric datasets, among others incorporating narrow-band H$\alpha$ imaging, have also enabled the search for symbiotic stars beyond the Milky Way. Targeted searches have identified candidates in M31 \citep[][]{2014MNRAS.444..586M}, M33 \citep[][]{2017MNRAS.465.1699M}, or LMC \citep{2018arXiv181106696I}. Additionally, symbiotic stars and candidates have been discovered as by-products of emission-line surveys focused on other populations, leading to detections in external galaxies such as IC 10 \citep[][]{2008MNRAS.391L..84G}, NGC 6822 \citep[][]{2009MNRAS.395.1121K,2015A&A...574A.102S}, NGC 185 \citep[][]{2012MNRAS.419..854G}, NGC 205 \citep[][]{2015MNRAS.447..993G}, and NGC 55 \citep[][]{2017MNRAS.464..739M}.

It is important to note that many of the aforementioned searches imposed relatively high detection thresholds for H$\alpha$ emission, meaning that even genuine H$\alpha$ emitters with weaker emission may have been overlooked (see panel D in Fig. \ref{fig:spectra}). As a result, there are still undetected symbiotic stars awaiting discovery. A recent example is Y Gem, a system known to be peculiar for years due to its association with X-ray emission from an AGB star, yet its symbiotic nature was only recently confirmed through spectroscopic follow-up \citep[][]{2025A&A...693A.203G}. {Moreover, surveys such as IPHAS, SHS, and VPHAS+ cover only a limited portion of the sky, typically within $|b| < 5$\textdegree --10\textdegree, leaving large areas, particularly at higher Galactic latitudes, unexplored. This limitation highlights the need for complementary search strategies, as discussed in the following subsection.}

{To conclude this section, one should add that} recent technological advancements now allow for the development of custom-designed narrow-band filters targeting specific emission lines beyond common H$\alpha$ or [O\,{\sc iii}]. A notable example in the search for symbiotic stars is the Raman Search for Extragalactic Symbiotic Stars Project \citep[RAMSES;][]{2019AJ....157..156A}, which utilizes a narrow-band filter centered on the Raman O\,{\sc vi} emission line. As already mentioned, this feature is unique to symbiotic binaries and is observed in approximately half of known systems \citep[][]{2019RNAAS...3...28M,2019AN....340..598M,2019ApJS..240...21A}. By selecting sources with excess emission in this filter, combined with detections in He\,{\sc ii} filters, RAMSES efficiently identifies strong symbiotic candidates for spectroscopic confirmation.

\subsection{Alternative photometric candidate selection methods}

The advent of new ground- and space-based datasets has enabled alternative approaches to identifying symbiotic stars beyond traditional searches based on strong emission lines. This is particularly relevant for systems with weak emission lines, which may occur if the white dwarf does not sustain shell burning, resulting in a lower temperature and luminosity, or if the accretor is a neutron star. These accreting-only symbiotic stars are more challenging to detect in the optical domain but can often be identified through UV excesses, flickering variability, and/or X-ray emission \citep[see, e.g.,][]{2013A&A...559A...6L,2016MNRAS.461L...1M}.

\citet{2023MNRAS.519.6044A} utilized a multi-wavelength approach by combining ultraviolet data from the Galaxy Evolution Explorer \citep[GALEX;][]{2017ApJS..230...24B} with infrared photometry from the Two Micron All Sky Survey \citep[2MASS;][]{2006AJ....131.1163S} and the Wide-field Infrared Survey Explorer \citep[WISE;][]{2010AJ....140.1868W}. Additionally, variability information from the All-Sky Automated Survey for Supernovae \citep[ASAS-SN;][]{2014ApJ...788...48S} was incorporated to refine candidate selection. GALEX UV photometry, combined with \textit{Gaia} DR3 data \citep[][]{2023A&A...674A...1G}, was also employed by \citet{2024ApJ...962..126X} to search for symbiotic stars.

A similar approach was taken by \citet{2024arXiv241200855L}, who also leveraged the fact that symbiotic binaries should exhibit a UV excess compared to single giants. They selected candidates based on colors from the SkyMapper Southern Sky Survey \citep{2018PASA...35...10W,2019PASA...36...33O}. Additionally, they utilized the multiple observations of the same target within short time intervals to identify minute-scale variability indicative of flickering, a signature of accretion discs.

Significant potential also lies in ongoing and future optical photometric surveys that have yet to be fully explored for symbiotic stars. These include the Javalambre-Photometric Local Universe Survey \citep[J-PLUS;][]{2019A&A...622A.176C} and the Southern Photometric Local Universe Survey\footnote{An effort to identify H$\alpha$ emitters in these data is ongoing, see, e.g., \citet{2025A&A...695A.104G}.} \citep[S-PLUS;][]{2019MNRAS.489..241M}, both of which observe the sky using 12 broad, intermediate, and narrow-band filters, the Javalambre Physics of the Accelerating Universe Astrophysical Survey \citep[J-PAS;][]{2014arXiv1403.5237B}, with its 54 narrow-band filters supplemented by intermediate and broad bands, or the Vera C. Rubin Observatory and its Legacy Survey of Space and Time (LSST) that will provide deep imaging in five intermediate-band filters \citep[][]{2022ApJS..258....1B,2023PASP..135j5002H}. Upcoming UV missions as the Ultraviolet Explorer \citep[UVEX;][]{2021arXiv211115608K}, the Ultraviolet Transient Astronomy Satellite \citep[ULTRASAT;][]{2024ApJ...964...74S}, and the Quick Ultra-VIolet Kilonova surveyor \citep[QUVIK;][]{2024SSRv..220...11W,2024SSRv..220...24K,2024SSRv..220...29Z}, are particularly relevant for searches based on UV excess.

\subsection{Variability-based search}
As discussed earlier, some symbiotic stars have been discovered due to their prominent brightness changes. Even today, previously unnoticed stars are sometimes classified as symbiotic binaries after exhibiting outbursts of several magnitudes. Recent examples include Hen 3-860, which was detected in outburst by the ASAS-SN survey \citep[][]{2022MNRAS.510.1404M}, as well as Gaia18aen \citep[][]{2020A&A...644A..49M} and Gaia23ckh \citep[=V390 Sco;][]{2024AN....34540017M}, whose brightenings were detected by the \textit{Gaia} satellite and reported as part of the \textit{Gaia} Photometric Science Alerts \citep[][]{2021A&A...652A..76H}. However, it is highly likely that \textit{Gaia} and similar transient photometric surveys have detected brightenings of other, yet unknown symbiotic stars, but these transients remain unclassified. Unfortunately, a large fraction of these events are never followed up spectroscopically due to the limited observational resources available. This challenge, relevant not only for symbiotic stars but the entire transient community, will become even more pronounced with upcoming large-scale surveys such as LSST, which will generate orders of magnitude more alerts per night, many of them from faint sources.

Additionally, some brightenings detected by surveys may never have been alerted publicly. For instance, the \textit{Gaia} Photometric Science Alerts system deliberately avoided triggering alerts on particularly red stars to prevent frequent alerts every time some Mira-type variable reaches maximum brightness. The current availability of the long-term light curves for a large number of stars is therefore tempting to search for additional symbiotic stars. However, one needs to keep in mind that symbiotic stars exhibit various forms of variability beyond outbursts, related to the orbital motion or the components themselves, like the pulsations of the giant (see Fig. \ref{fig:lcs}). In \textit{Gaia} DR3, machine-learning techniques were applied to classify variable stars, including symbiotic binaries \citep[][]{2023A&A...674A..13E,2023A&A...674A..14R}. However, due to the similarities between the quiescent light curves of symbiotic stars and those of pulsating giants and the shared position of these two groups in the HR diagram, the resulting symbiotic sample is significantly contaminated by single pulsating giants. Additional information, such as some binary indicator or the presence of emission lines, would be necessary to distinguish these populations, but such criteria were not included in the \textit{Gaia} DR3 classification algorithm. The upcoming Gaia DR4, which will provide astrometric, photometric, and spectroscopic time-series data, holds therefore  great promise for uncovering many new symbiotic systems.

\subsection{Searching in spectroscopic datasets}

In recent years, alongside the large photometric catalogs (having time series data or not), massive spectroscopic surveys have been producing an unprecedented amount of data, e.g., the Large Sky Area Multi-Object Fiber Spectroscopic Telescope \citep[LAMOST;][]{2012RAA....12.1197C}, the GALactic Archaeology with HERMES survey \citep[GALAH;][]{2015MNRAS.449.2604D,2021MNRAS.506..150B}, the Sloan Digital Sky Survey \citep[SDSS V;][]{2017arXiv171103234K}, or the Dark Energy Spectroscopic Instrument \citep[DESI;][]{2016arXiv161100036D}. In addition, several upcoming spectroscopic surveys promise similarly or even more extensive datasets, some starting soon, e.g., 4-metre Multi-Object Spectroscopic Telescope \citep[4MOST;][]{2019Msngr.175....3D}, or William Herschel Telescope Enhanced Area Velocity Explorer \citep[WEAVE;][]{2024MNRAS.530.2688J}, some are planned in the more distant future like the Wide-Field Spectroscopic Telescope \citep[WST;][]{2024arXiv240305398M}.

These large spectroscopic datasets provide an additional unique opportunity to search for symbiotic stars. \citet[][]{2021MNRAS.505.6121M} utilized data from the GALAH survey to specifically search for accreting-only symbiotic stars. Their approach focused on identifying M giant stars exhibiting weak H$\alpha$ emission while filtering out cases where the emission originates from internal shock regions in pulsating giants.

\subsection{A few concluding words for the search of symbiotic stars}
Several words of caution are warranted when considering the search for symbiotic stars. Many of the approaches discussed above have yielded numerous candidates; however, only a fraction have been followed up spectroscopically. Until confirmed by spectroscopy, these objects cannot be regarded as genuine symbiotic stars, regardless of how well they fit the selection criteria. Unfortunately, it is not uncommon for unconfirmed candidates to be prematurely classified as bona-fide symbiotic stars in databases such as SIMBAD or VSX. These misclassifications can propagate errors into studies on symbiotic populations or machine-learning training samples. Follow-up observations often reveal that many of these candidates do not, in fact, belong to the symbiotic class.

Particular caution is needed when the selection process relies on somehow limited data. For instance, as mentioned earlier, the \textit{Gaia} DR3 variability classification struggles to distinguish between symbiotic stars and pulsating giants due to data (astrometry, colors, variability) that were employed for this classification. An additional example is the work of \citet[][]{2023RAA....23j5012J} who employed machine-learning techniques to search for symbiotic stars among LAMOST DR9 sources, using only infrared photometry from 2MASS and WISE. While their method identified thousands of candidates, they only examined SDSS spectra for 15 of them. Two of these showed emission lines observed, among other objects, also in symbiotic stars and were therefore classified by the authors as newly discovered symbiotics. However, both were previously already known as young stellar objects, and they clearly cannot host red giants, given their distances and apparent magnitudes. Their classification algorithm did not incorporate astrometric data or photometry beyond the infrared, though. 

X-ray surveys have also yielded symbiotic candidates, often based solely on the positional coincidence of a bright X-ray source with a red object exhibiting symbiotic-like 2MASS/WISE colors \citep[e.g.,][]{2020MNRAS.499.3111S,2022MNRAS.512.5481S,2022A&A...661A..38P}. However, such samples can be contaminated by other sources, such as reddened early-type stars or chromospherically active close binaries (e.g., RS CVn-type systems). Thus, additional spectroscopic confirmation is always necessary. An interesting case in this regard is the discovery of a symbiotic candidate in the globular cluster 47 Tuc \citep[][]{2022A&A...661A..35S}. Symbiotic stars are predicted to be extremely rare in globular clusters, and if present, they should be located far from the cluster core \citep[][]{2020MNRAS.496.3436B}. A similar case of a symbiotic candidate in $\omega$~Cen was ultimately shown to be a misclassification, resulting from a chance alignment of a carbon star and an unrelated X-ray source \citep[][]{2020MNRAS.496.3436B}.

Even when comprehensive survey data and optical spectroscopy are available, definitive classification is sometimes still elusive, especially when emission lines with high ionization potential are absent (see panels D and E of Fig. \ref{fig:spectra}). In such cases, observations at other wavelengths become crucial. UV spectroscopy, for example, has historically been used to confirm the symbiotic nature of certain stars using data from the International Ultraviolet Explorer \citep[IUE;][]{1978Natur.275..372B} and the Far Ultraviolet Spectroscopic Explorer \citep[FUSE;][]{2000ApJ...538L...1M}. However, the availability of UV spectra is nowadays quite limited, and only a handful of symbiotic stars have been studied in this way in recent years. A recent example is the observation of SU Lyn using the ASTROSAT mission of Indian Space Research Organisation \citep[][]{2014SPIE.9144E..1SS}, presented by \citet{2021MNRAS.500L..12K}. UVEX, planned for launch in 2030, would have spectroscopic capabilities as well \citep[][]{2021arXiv211115608K}. 

The use of large datasets also necessitates rigorous validation. Data reduction and processing techniques applied to extensive datasets can sometimes lead to spurious results for some individual objects whose observations are affected in unexpected ways and not accounted for in the automated reduction pipelines (often, the results are not validated on the object-by-object basis). A notable example is LAMOST J202629.80+423652.0, initially classified as a symbiotic star by \citet{2015RAA....15.1332L} due to the apparent presence of emission lines in its LAMOST spectrum. However, subsequent analyses revealed that the emission lines were not real but rather an artifact of inaccurate sky-background subtraction \citep[][]{2020CoSka..50..672A}.

On a more optimistic note, many of the ground- and space-based surveys discussed earlier (and in the following sections) can be leveraged not only for identifying new candidates but also for re-evaluating previously identified ones. New data sometimes allow the confirmation of the symbiotic status, while some stars originally classified as symbiotics have been found to lack red giants altogether \citep[e.g.,][]{2020MNRAS.499.2116M,2021MNRAS.506.4151M}. For example, many of the symbiotic candidates selected based on H$\alpha$ photometry have since been shown to be young stellar objects. The integration of multi-wavelength data and follow-up observations will remain crucial for refining symbiotic star catalogs and ensuring accurate classification.

\section{Understanding the symbiotic variability}
\begin{figure}[t]
\isPreprints{\centering}{} % Only used for preprints
\includegraphics[width=13.4 cm]{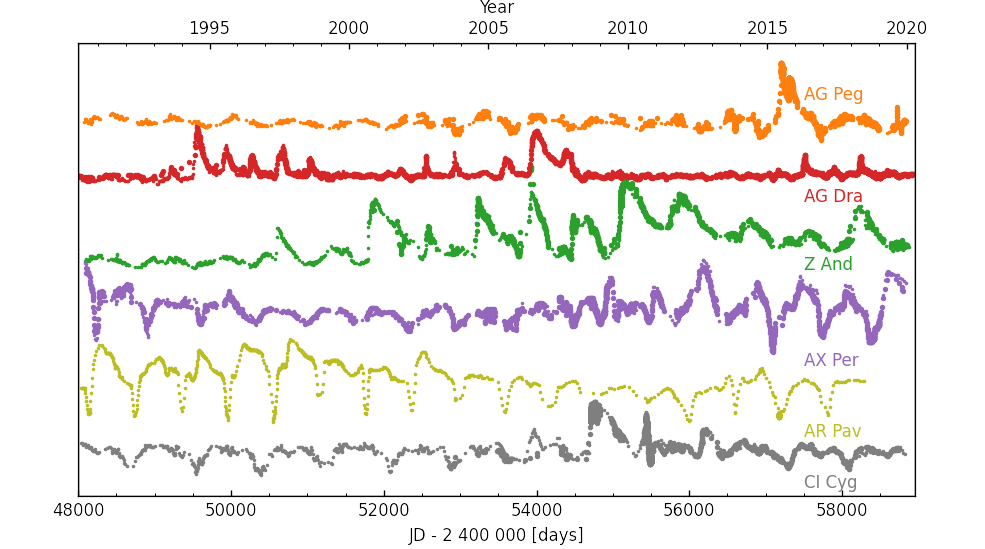}
\caption{Long-term light curves of selected symbiotic stars spanning 30 years. Variability associated with outburst activity and orbital motion is evident. Smaller dots represent visual observations, while larger dots correspond to CCD measurements in the $V$ filter. A running average has been applied to the visual data for clarity. The figure adapted from \citep{2022PhDT........34M}. The data are primarily sourced from the American Association of Variable Star Observers (AAVSO) database \citep[for details, see][]{2022PhDT........34M}.\label{fig:lcs}}
\end{figure}

The orbital periods of known symbiotic systems span from several hundred to thousands of days. Excluding the exceptionally long orbital periods of D-type symbiotics, systems that host highly evolved, large red giants that typically pulsate as Miras, the more common S-type symbiotic stars exhibit orbital period distributions peaking between 500 and 600 days. The red giants in symbiotic systems typically undergo semi-regular pulsations with periods ranging from approximately 40 to 200 days, while Mira-type pulsators exhibit longer pulsation periods, often spanning several hundred days \citep[e.g.,][]{2009AcA....59..169G,2013AcA....63..405G,2019RNAAS...3...28M}. Furthermore, symbiotic binaries can undergo prolonged episodes of activity, including recurrent Z And-type outbursts and "slow" symbiotic nova eruptions, which may persist for decades. Given these characteristic timescales, analyzing the variability of symbiotic stars requires extensive long-term time-series observations (see~Fig.~\ref{fig:lcs}).

\subsection{Symbiotics in the long-term photometric surveys}
Several past and ongoing photometric surveys have provided valuable time-series data for studying symbiotic stars. These include, among others, the previously mentioned ASAS-SN, as well as the Northern Sky Variability Survey \citep[NSVS;][]{2004AJ....127.2436W}, the All-Sky Automated Survey \citep[ASAS;][]{1997AcA....47..467P}, the Super Wide Angle Search for Planets \citep[SuperWASP;][]{2010A&A...520L..10B}, the Optical Gravitational Lensing Experiment \citep[OGLE;][]{2008AcA....58...69U,2015AcA....65....1U}, the Asteroid Terrestrial-impact Last Alert System \citep[ATLAS;][]{2018PASP..130f4505T,2020PASP..132h5002S}, and the Zwicky Transient Facility \citep[ZTF;][]{2019PASP..131a8003M}. While the primary goals of these surveys often differ from the direct study of stellar variability, their extensive datasets have proven highly useful in this field, leading to the discovery of numerous unexpected and peculiar variable stars.

These survey data are now routinely employed in the analysis of both newly identified and previously known symbiotic stars \citep[for recent examples, see, e.g.,][]{2022MNRAS.510.1404M,2021MNRAS.506.4151M,2020MNRAS.499.2116M,2020A&A...636A..77S,2020A&A...644A..49M,2023MNRAS.526.6381S,2023ApJ...953L...7I,2024PASP..136g4202N,2024AN....34540017M,2025A&A...693A.203G}. Long-term light curves have also been instrumental in studying symbiotic variability as a population, as demonstrated by \citet{2009AcA....59..169G,2013AcA....63..405G}, who analyzed light curves from surveys such as ASAS and OGLE. Their work provided numerous new orbital and pulsation periods, helping to create a less biased sample of symbiotic orbital periods, which is crucial for comparing observed populations with theoretical predictions from population synthesis studies.

While this discussion focuses on modern surveys, advances in technology have also enabled the digitization and analysis of historical photographic plate data. The Digital Access to a Sky Century at Harvard (DASCH) archive \citep{2010AJ....140.1062L} is one such resource, allowing researchers to examine past brightness variations of symbiotic stars and analyze previous outbursts \citep[e.g.,][]{2015MNRAS.451.3909I,2016IBVS.6176....1M,2019A&A...624A.133I,2020A&A...636A..77S,2022MNRAS.510.1404M}, including the symbiotic recurrent novae such as RS Oph \citep[][]{2010ApJS..187..275S}, T CrB \citep[][]{2020ApJ...902L..14L,2023MNRAS.524.3146S,2024MNRAS.532.1421T}, and V745 Sco \citep[][]{2024RNAAS...8...93S}.

These datasets are also invaluable when an unknown star is detected in outburst or identified as a potential symbiotic system. They enable researchers to analyze the variability of these objects without requiring long-term targeted follow-up. However, there are certain limitations to consider. Each survey has specific brightness limits (e.g., ASAS-SN is particularly useful for brighter stars, while ZTF and OGLE provide better coverage for fainter sources). In some cases, a star may have usable data in either quiescence or outburst but not in both, due to its variability amplitude exceeding the detection limits of the survey. Other challenges include differences in angular resolution (surveys like ASAS-SN have relatively low resolution, leading to blending issues in crowded fields). Additionally, some surveys, such as OGLE, cover only specific sky regions, limiting their availability for certain targets. Finally, most surveys provide data in only one filter (or in unfiltered mode), meaning that color information, which can be critical for understanding the physical causes of variability, is often unavailable.

\subsection{Short timescales from space}
The surveys discussed in the previous section typically provide observations with a cadence of a few days. However, much higher cadence is required to study the shortest variability timescales in symbiotic stars, such as flickering from accretion discs \citep[][]{1996AJ....111..414D,2001MNRAS.326..553S,2006AcA....56...97G} or the rotation of magnetic white dwarfs in some systems \citep[][]{1999ApJ...517..919S}. This is now possible thanks to data from space-based missions such as the \textit{Kepler} space telescope \citep[][]{2010Sci...327..977B}, the Transiting Exoplanet Survey Satellite \citep[\textit{TESS};][]{2015JATIS...1a4003R}, and, in the near future, the PLAnetary Transits and Oscillations of Stars satellite \citep[\textit{PLATO};][]{2024arXiv240605447R}.

While these space-based datasets have certain limitations for studying symbiotic stars, such as observations in red passbands where variability from hot component is less pronounced, the large pixel scale of \textit{TESS}, and limited target selection, they have already been successfully used to investigate individual systems, including CN Cha \citep{2020AJ....160..125L}, T CrB \citep{2023MNRAS.524.3146S}, RT Cru \citep{2023A&A...670A..32P}, and IGR J16194-2810 \citep{2023A&A...676L...2L,2024PASP..136g4202N}. More recently, \citet{2024A&A...689A..86L} analyzed the \textit{TESS} light curves of V1261 Ori, NQ Gem, and V420 Hya. Additionally, \textit{TESS} data for a large sample of symbiotic stars were utilized by \citet{2024A&A...683A..84M} to search for new flickering sources. This study confirmed flickering-like variability in 13 new sources, adding to 20 previously know. Now, flickering has been detected in approximately 80\% of accreting-only symbiotic stars, reinforcing the idea that accretion disks are common in these binaries.

High-cadence observations from \textit{TESS} and \textit{Kepler} have also proven useful for studying short-period periodic signals in symbiotic systems, which are thought to be associated with the rotation of magnetic white dwarfs. Such signals were previously detected in CH Cyg \citep[8.3 min;][]{1990AcA....40..129M}, Z And \citep[28 min;][]{1999ApJ...517..919S,2024A&A...683A..84M}, BF Cyg \citep[108 min;][]{2009MNRAS.396.1507F}, and Hen 2-357 \citep[31.4 min; ][]{2016MNRAS.463.1099T}. More recently, \citet{2023A&A...675A.140M} identified an 11.3-minute periodicity in the \textit{Kepler} light curves of FN Sgr, while \citet{2024A&A...683A..84M} reported similar signals in \textit{TESS} observations of AE Cir (27.3 min) and possibly CI Cyg (66.6 min), though further confirmation is needed. Notably, the period of Z And appears to have changed by approximately 80 seconds over the 25 years since its initial detection, with \textit{TESS} measuring a period of 26.7 minutes \citep{2024A&A...683A..84M}.

Recent studies suggest that the late emergence of magnetic fields in white dwarfs \citep[][]{2021MNRAS.507.5902B,2021MNRAS.502.4305P} can significantly impact their evolution \citep[][]{2021NatAs...5..648S}. However, the mechanisms responsible for generating this magnetism remain poorly understood. A study of FN Sgr by \citet{2024A&A...686A.226B} proposed that symbiotic stars could be key to understanding magnetism in close double white dwarf binaries. Expanding the sample of known magnetic symbiotic systems is therefore essential for constraining models of magnetic white dwarf formation and evolution.

Finally, it is worth noting that \textit{Kepler} K2 and \textit{TESS} data have also been used to investigate compact objects in symbiotic X-ray binaries. \citet[][]{2023A&A...676L...2L} studied the spin-down of the pulsar in GX 1+4 and also detected the optical spin period of the neutron star in the symbiotic X-ray binary IGR J16194-2810.

\subsection{Time-series radial velocities}

To complete the discussion of variability, it is worth noting that time-series radial velocity measurements are gradually becoming available from some large-scale surveys, such as the Radial Velocity Experiment \citep[RAVE;][]{2006AJ....132.1645S} and the Apache Point Observatory Galactic Evolution Experiment \citep[APOGEE;][]{2017AJ....154...94M}. Notably, APOGEE data enabled the first spectroscopic orbit determination of a symbiotic star beyond the Milky Way, Draco C1 in the Draco Dwarf galaxy \citep[][]{2020ApJ...900L..43L}. However, the vast majority of spectroscopic orbits for symbiotic stars are still obtained through dedicated, long-term follow-up campaigns.

A small number of orbital solutions for symbiotic stars were also published as part of \textit{Gaia} DR3 \citep[][]{2023A&A...674A..34G,2024A&A...682A...7B}, along with radial velocity measurements for a few symbiotic systems included in the \textit{Gaia} Focused Product Release \citep[][]{2023A&A...680A..36G}. However, the timespan of \textit{Gaia} DR3 is still relatively short for deriving fully reliable orbital solutions for most symbiotic stars, see, for example, the case of V1261 Ori discussed in \citet{2025A&A...695A..61M}. While the longer timespan of \textit{Gaia} DR4, combined with the availability of full time-series data, will improve the situation, radial velocities will likely remain most useful for the brightest systems.

\section{Beyond the surveys}
It is impossible to cover all recent advances in the field of symbiotic stars within this review. Instead, I have so far primarily focused on the impact of modern large-scale ground- and space-based surveys, which, as demonstrated in previous sections, have provided a wealth of valuable data. However, many critical research questions cannot yet be addressed solely with survey data and still require dedicated follow-up observations.

For instance, spectroscopic orbit determinations remain largely dependent on targeted long-term monitoring, see, for example, the series of papers on spectroscopic orbits by Fekel, Hinkle, and collaborators \citep[e.g.,][]{2017AJ....153...35F, 2019ApJ...872...43H}. Similarly, detailed analyses of the chemical abundances of the giant components in symbiotic systems require high-resolution spectroscopy \citep[e.g.,][and references therein]{2023MNRAS.526..918G,2017MNRAS.466.2194G}. Spectroscopic follow-up is also crucial for confirming the classification of newly identified candidates and for tracking the spectral evolution of known symbiotic stars, particularly during their outbursts. Despite the increasing availability of all-sky photometric surveys, dedicated photometric monitoring remains essential for fully characterizing some objects.

Moreover, certain types of observations will likely never be routinely obtained through large-scale surveys, for example interferometric observations, which have provided unique insights into symbiotic stars \citep[see, e.g.,][]{2025A&A...695A..61M,2014A&A...564A...1B}. Targeted deep imaging, reaching significantly fainter limits or achieving much higher angular resolution than surveys, has also yielded interesting discoveries. Noteworthy examples include the identification of ancient nova shells around RX Pup, suggesting past variations in its mass transfer rate \citep[][]{2024ApJ...972L..14I}, and the detection of an extended nova super-remnant surrounding the recurrent symbiotic nova T CrB \citep[][]{2024ApJ...977L..48S}. A similar structure may yet be discovered around RS Oph \citep[][]{2024MNRAS.529L.175H}. Additionally, long-term monitoring of R Aqr has enabled detailed studies of the evolution of its jets over time \citep[e.g.,][]{2021gacv.workE..41L}. 

\subsection{Contribution of amateur astronomers}

It is essential to highlight the significant contributions of amateur astronomers to dedicated photometric and spectroscopic follow-up observations \citep[see also][and references therein]{2021OEJV..220...11M}. Equipped with relatively small telescopes, amateur observers play an important role in monitoring symbiotic stars. Their photometric data, typically archived in databases such as those of the American Association of Variable Star Observers (AAVSO) and the British Astronomical Association (BAA), and their spectroscopic observations, available in repositories like the Astronomical Ring for Amateur Spectroscopy \citep[ARAS;][]{2019CoSka..49..217T} or AAVSO, have been widely used in professional research. In many cases, these data have been indispensable for the studies \citep[for examples, see, e.g.,][]{2021MNRAS.506.4151M,2020A&A...636A..77S,2022MNRAS.510.1404M,2020MNRAS.492.3107L,2022MNRAS.515.4655P,2022MNRAS.510.2707I}.

Amateur astronomers have also contributed to the discovery of new symbiotic stars. Notable examples include DeGaPe 35, identified through an amateur-led survey of planetary nebula candidates \citep[][]{2023NewA...9801943P}, or TCP J18224935-2408280, a new symbiotic star detected in outburst by an amateur observer \citep[][]{2021ATel14691....1M,2021ATel14699....1T,2021ATel14692....1A,2023MNRAS.526.6381S}. These results underscore the valuable synergy between professional and amateur astronomers in the study of symbiotic stars.

\subsection{Going beyond the UV, optical and infrared}
The contribution of X-ray observations to the search for symbiotic stars, both in the Milky Way and in Local Group galaxies, has already been highlighted. Some systems were identified due to the positional coincidence of an X-ray source with a red giant, while others, such as SRGA J181414.6-225604, were selected for more detailed follow-up because of their high X-ray variability \citep[][]{2022ApJ...935...36D}. Currently, a relatively shallow all-sky X-ray survey is available thanks to eROSITA (the extended ROentgen Survey with an Imaging Telescope Array) mission \citep[][]{2021A&A...647A...1P, 2024A&A...682A..34M}, with half of the sky accessible to the international community. However, most other X-ray observatories, such as the Chandra X-ray Observatory \citep[][]{2000SPIE.4012....2W}, XMM-Newton \citep{2001A&A...365L...1J}, NuSTAR \citep[the Nuclear Spectroscopic Telescope Array;][]{2013ApJ...770..103H}, or Swift \citep[][]{2004ApJ...611.1005G}, have observed only small portions of the sky. Therefore, targeted X-ray observations remain essential for the detailed study of symbiotic stars.

At the same time, increasingly higher spectral resolution observations are becoming available, providing unprecedented insight into the physics of accretion and outbursts. For example, the symbiotic recurrent novae V745 Sco, RS Oph, and V3890 Sgr have been extensively monitored in X-rays (as well as across basically all other wavelength domains) during their recent outbursts \citep[see, e.g.,][]{2015MNRAS.454.3108P, 2020MNRAS.499.4814P, 2022A&A...658A.169N, 2022MNRAS.514.1557P, 2023A&A...670A.131N, 2022ApJ...938...34O, 2023ApJ...955...37O}. Many symbiotic systems have been studied in detail in X-rays for the first time, thanks to the increasing availability of observational data \citep[see, e.g.,][]{2024MNRAS.527.3585B, 2024A&A...689A..86L, 2023ApJ...948...14T}.

Long-term X-ray monitoring has revealed dramatic changes in some systems. In the case of RT Cru, for example, variations in the X-ray spectrum and luminosity were linked to changes in the accretion rate \citep[][]{2023A&A...670A..32P}. R Aqr displayed changes in both its X-ray and optical properties during periastron passage \citep{2024MNRAS.535.2724V}. \citet{2022ApJ...927L..20T} studied the extended X-ray emission in this system spatially correlated with its optical nebula.

Recent studies suggest that switching between different types of X-ray emission may be driven by reflection effects and variations in disk properties \citep[][]{2023MNRAS.522.6102T, 2024MNRAS.528..987T}, raising questions about the sufficiency of the previously used X-ray classification scheme \citep[][and references therein]{2013A&A...559A...6L}. Additionally, X-ray observations of T CrB prior to its anticipated eruption have revealed an evolution pattern in the hardness-intensity diagram reminiscent of those seen in black hole binaries, accreting neutron stars, and active galactic nuclei \citep[][]{2024MNRAS.532.1421T}. This further underscores the broader astrophysical relevance of symbiotic binaries, demonstrating their connections to other high-energy astrophysical phenomena.

With the advent of new observational facilities, significant progress has been made at both very long (radio) and very short ($\gamma$-ray) wavelengths. Radio observations have played a crucial role in studying recent outbursts of the aforementioned symbiotic recurrent novae, enabling high-resolution imaging of evolving outflows and providing unique insights into the physics of these eruptions \citep[][]{2023MNRAS.523.1661N, 2024A&A...692A.107L, 2024MNRAS.534.1227M, 2024MNRAS.528.5528N}.

At the opposite end of the spectrum, several symbiotic novae have been detected in $\gamma$-rays, including V407 Cyg \citep[][]{2010Sci...329..817A}, RS Oph \citep[][]{2022ApJ...935...44C}, and V3890 Sgr \citep[][]{2019ATel13114....1B}. Notably, the 2021 outburst of RS Oph marked a significant milestone as the first nova eruption ever detected in very-high-energy ($>$100 GeV) $\gamma$-rays \citep[][]{2022Sci...376...77H, 2022NatAs...6..689A}.

\subsection{Theoretical perspectives}

While this review has primarily focused on observational findings, significant progress is also being made on the theoretical front, advancing our understanding of symbiotic stars. Notably, recent studies on mass transfer, such as those by \citet{2024ApJ...971...64I}, \citet{2025ApJ...980..226T}, \citet{2025arXiv250211325M}, and \citet{2025ApJ...980..224V}, provide valuable insights into the accretion processes in these systems. A proper understanding of mass transfer is crucial for detailed evolutionary modeling, as demonstrated, for example, in the case of FN Sgr by \citet{2024A&A...686A.226B}. Ultimately, such advancements pave the way for comprehensive population synthesis studies of symbiotic stars. So far, significant discrepancies remain between the observed properties of symbiotic stars and predictions from population studies \citep[e.g.,][]{2006MNRAS.372.1389L}. The role of symbiotic systems as potential progenitors of Type Ia supernovae continues to be an area of active investigation as well, see, for example, the works by \citet{2022ApJ...941L..33A}, \citet{2023RAA....23g5010L}, \citet{2021MNRAS.503.4061W}, \citet{2019MNRAS.485.5468I} {and \citet{2025arXiv250402090L}. While there is clear observational evidence that white dwarfs in symbiotic systems can grow in mass \citep{2017ApJ...847...99M}, and theoretical studies suggest that interactions even in some of the widest D-type symbiotic binaries could lead to Type Ia supernovae \citep{2019MNRAS.485.5468I}, current findings indicate that symbiotic stars are unlikely to represent the dominant population of progenitors, at least within the single-degenerate channel \citep[e.g.,][]{2025arXiv250402090L}. The most promising candidates are systems exhibiting recurrent nova outbursts, such as T~CrB or RS~Oph, which are known to host very massive white dwarfs \citep{2017ApJ...847...99M,2021MNRAS.504.2122M,2025ApJ...983...76H}. However, only a handful of such objects are known. It is likely that more such systems exist, but they are observationally challenging to identify in quiescence, as in the accreting state, they can closely resemble ordinary single giants. }

To conclude, even in the era of modern observational surveys, advanced computational methods, and powerful supercomputers, the evolution and ultimate fate of symbiotic stars remain only partially understood. The prospects for future research are, however, highly promising, with new discoveries and theoretical advancements continually reshaping our understanding of these interacting systems.

%%%%%%%%%%%%%%%%%%%%%%%%%%%%%%%%%%%%%%%%%%
%\section{Conclusions}

%This section is not mandatory, but can be added to the manuscript if the discussion is unusually long or complex.

%%%%%%%%%%%%%%%%%%%%%%%%%%%%%%%%%%%%%%%%%%
%\section{Patents}

%This section is not mandatory, but may be added if there are patents resulting from the work reported in this manuscript.

%%%%%%%%%%%%%%%%%%%%%%%%%%%%%%%%%%%%%%%%%%
\vspace{6pt}

\funding{J.M. was supported by the Czech Science Foundation (GACR) project no. 24-10608O and by the Spanish Ministry of Science and Innovation with the grant no. PID2023-146453NB-100 (PLAtoSOnG).}

\dataavailability{No new data were created in the scope of this work.}

\acknowledgments{The author thanks both anonymous reviewers for their constructive comments and suggestions, which helped improve the clarity and quality of the manuscript. The author expresses gratitude to the organizers of the "Hot Stars—Life with Circumstellar Matter" conference for the invitation and support to attend the meeting.}

\conflictsofinterest{The author declares no conflicts of interest.} 

\isPreprints{}{% This command is only used for ``preprints''.
\begin{adjustwidth}{-\extralength}{0cm}
} % If the paper is ``preprints'', please uncomment this parenthesis.

\reftitle{References}


\bibliography{bibliography.bib}

% If authors have biography, please use the format below
%\section*{Short Biography of Authors}
%\bio
%{\raisebox{-0.35cm}{\includegraphics[width=3.5cm,height=5.3cm,clip,keepaspectratio]{Definitions/author1.pdf}}}
%{\textbf{Firstname Lastname} Biography of first author}
%
%\bio
%{\raisebox{-0.35cm}{\includegraphics[width=3.5cm,height=5.3cm,clip,keepaspectratio]{Definitions/author2.jpg}}}
%{\textbf{Firstname Lastname} Biography of second author}

% For the MDPI journals use author-date citation, please follow the formatting guidelines on http://www.mdpi.com/authors/references
% To cite two works by the same author: \citeauthor{ref-journal-1a} (\citeyear{ref-journal-1a}, \citeyear{ref-journal-1b}). This produces: Whittaker (1967, 1975)
% To cite two works by the same author with specific pages: \citeauthor{ref-journal-3a} (\citeyear{ref-journal-3a}, p. 328; \citeyear{ref-journal-3b}, p.475). This produces: Wong (1999, p. 328; 2000, p. 475)

%%%%%%%%%%%%%%%%%%%%%%%%%%%%%%%%%%%%%%%%%%
%% for journal Sci
%\reviewreports{\\
%Reviewer 1 comments and authors’ response\\
%Reviewer 2 comments and authors’ response\\
%Reviewer 3 comments and authors’ response
%}
%%%%%%%%%%%%%%%%%%%%%%%%%%%%%%%%%%%%%%%%%%
\PublishersNote{}
\isPreprints{}{% This command is only used for ``preprints''.
\end{adjustwidth}
} % If the paper is ``preprints'', please uncomment this parenthesis.
\end{document}